| | |
|---|---|
| Abstract | Simulations and experimental works have been carried out in a complementary way to engineer a basic material target mimicking the same dielectric properties of the human body. It includes a resistor in parallel with a capacitor, whose values ($R_h$=1500 Ω and $C_h$=100 pF) are estimated in regard of parameters commonly utilized upon *in vivo* campaigns (frequency=30 kHz, gap=10 mm, high voltage electrode surface=12.6 mm²). This equivalent electrical human body (EEHB) circuit can be used as a reference and realistic target to calibrate electrical properties of therapeutic plasma sources before their utilization on patients. In this letter, we consider a configuration where this EEHB target interacts with a plasma gun (PG). Plasma power measurements performed in such configuration clearly indicate two operating modes depending on the value of the supplied voltage. Hence, the plasma gun generates pulsed atmospheric plasma streams likely to present therapeutic interest for voltages comprised between 3.0 and 8.5 kV while for higher values, transient arcs of thermal plasma are generated and represent substantial risks for the patient. |

# I. Introduction

Atmospheric pressure plasma jets (APPJ) are widely investigated as therapeutic options for medical applications (Laroussi 2015) (Weltmann and Woedtke 2017) (Park et al 2018). When they are supplied with nanosecond (or microsecond) high voltage pulses, these devices provide transient ionization waves of plasma called atmospheric-pressure plasma streams (PAPS), including the so-called plasma bullets phenomena (Robert et al 2012), with the ability to propagate on long distances in dielectric capillaries (from few mm to 1 m) (Sarron et al 2011). Physical properties of a plasma used for therapeutic purposes, depend on the source itself and on the target with which it interacts (Norberg et al 2015) (Kone et al 2017). For example, Darny et al have demonstrated how a grounded metallic target influence the properties of a discharge generated by a plasma gun, owing to the impedance mismatch between the remnant ionized channel (behind the ionization front) and the target (Darny et al 2017). Guaitella and Sobota have investigated the interaction of APPJ with dielectric target: in comparison with a free jet configuration, the interaction with a glass plate can make the jet length twice longer by enhancing locally the electric field and by promoting the accumulation of deposited charges (Guaitella and Sobota 2015). For these reasons, plasma source and target cannot be envisioned separately: they must be considered as a unified and interactive system. The term "target" must be taken in the broad sense, distinguishing on the one hand the material targets (typically dielectric or conductive plates brought to a floating potential or to the ground) and on the other hand living targets which correspond to biological models (cell cultures, mice, rats, pigs) or human patients. Using material targets is a cheap, fast and simple approach to evaluate the electrical, chemical, radiative and flow properties of a plasma source. It allows to benchmark various plasma sources or setup configurations and select the most promising for a dedicated medical application (Winter et al 2015) (Darny et al 2017). This approach can also be used as a preliminary step before achieving *in vivo* experiments: the plasma source is calibrated at physics lab to define operating conditions to follow afterwards at hospital.

Today, most of the material targets utilized in the Plasma Medicine Community are limited to conductive or dielectric plates which, by definition, are not representative of living organisms on the point of electrical view. In that respect, they are not relevant to respond one of the most important issues: patient safety in regard of electrical hazards like transient plasma arcs which can induce electrical injuries like extensive and deep burns of tissues, ventricular fibrillation and neurological effects [Koumbourlis 2002]. As an alternative, one could be tempted to carry out *in vivo* experiments on murine models so as to define an experimental operating window by changing all the relevant parameters of the "plasma source target" system. This preliminary step could hence guarantee that the patient exposed to plasma encounters no risk. If this approach seems attractive at first glance, it has two drawbacks: first it is expensive in terms of time and mice to sacrifice, and second it may provide unusable results. Indeed, mouse and human have levels of biological organization drastically different, as are the values of relative permittivity and electrical conductivity of their tissues, organs and bones, inducing *de facto* electrical impedances of their own. To be even more rigorous, this complex impedance is not only target-dependant but target and plasma source dependant. Therefore, a dedicated experimental operating window must be attributed to every "living organism - plasma source" system.

In a perspective of translational research, the most reliable alternative to guarantee the safety of a plasma source is to work on material targets that best reproduce the electrical properties of a living organism, here a human body. In this letter, we propose to demonstrate the benefit of studying these EEHB targets by using a plasma source already successfully applied on murine models in the field of oncology: the plasma gun (Brullé et al 2012). In the framework of translational research, its application to patients may indeed require adjustments, likely to be determined using an EEHB target. Equivalent electrical circuits of the plasma gun and of the human body target are discussed as well as the most accurate method to determine the relevant electrical parameters, i.e. plasma power and electrical power deposited in the target.

## II. From the plasma gun device to its equivalent electrical circuit

A global schematic view of the plasma gun (PG) is represented in Figure 1. Plasma is generated in a 100 mm long quartz tube, with inner and outer diameters of 4 mm and 8 mm respectively. It is supplied with helium (1000 sccm) and powered by a mono-polar square pulse high voltage generator (Spellman, SLM 10 kV 1200 W) coupled with a Smart HV Pulses Generator (RLC electronic, NanoGen1 10 kV). The Plasma Gun source is a 50 mm long inner high voltage electrode centered in the tube. A 10 mm wide grounded ring electrode is fixed on the outer quartz tube. Its two circular contours are equally distant from the extremity of the inner high voltage electrode.

Usually, the equivalent electrical circuit of a dielectric barrier discharge device (DBD) with/without gas flowing is represented by two electronic components in parallel : (i) a capacitor standing for the capacitance of the non-ionized gas (plasma off) confined in the electrode/dielectric/electrode sandwich structure and (ii) a variable plasma impedance $Z_{plasma}$ which can include a combination of other resistor and capacitor components depending on the model complexity (ion sheaths, space charge areas, boundary layers, etc.) (Kostov et al 2009) (Slutsker et al 2017). As sketched in Figure 1b, plasma in the interelectrode region is represented by $C_{IE}$ (capacitance of helium gas) in parallel with $Z_{bulk}$ while in the post-electrode region, i.e. between the counter electrode and the treated target, plasma is represented by $C_{PE}$ (capacitance of the helium-air gas mixture) in parallel with $Z_{plume}$. When plasma is off, $Z_{plume}$ tends to infinity (open circuit while the quartz tube behaves as a dielectric barrier wherever electrical charges are likely to accumulate. Two surfacic regions of the inner quartz tube are identified: where the coaxial and ring electrodes are axially facing together (region A in Figure 1a) and along the inner wall delimited by the outer ring electrode (region B in the Figure 1a). In these two regions, the quartz tube behaves as a dielectric barrier and is represented by wall capacitor: $C_{w1}$ and $C_{w2}$ respectively. A material target (electrically grounded or not) can be placed 10 mm downstream of the quartz tube to investigate how it can influence the discharge properties.

As described in Figure 1b, the high voltage electrode is connected to the generator through a $C_{m1}$ capacitor while the other is grounded through $C_{m2}$ (Murata 100 pF 6.3 kV DC). These capacitors allow the determination of $I_{gene}(t)$ and $I_{gr}(t)$ derived from equations {2} and {3} respectively from voltage or electrical charge from generator $Q_{gene}$ {1}. The instantaneous current of the plasma plume, i.e. of the ionization wave, is deduced from {4}.

Total power delivered by the generator ($P_{gene}$) and plasma power are calculated using high-voltage probes (Tektronix P6015A 1000:1, Teledyne LeCroy PPE 20 kV 1000:1, Teledyne LeCroy PP020 10:1) and an analog oscilloscope (HMO3004, Rohde & Schwarz). $P_{gene}$ is estimated by plotting a Lissajous curve according to {5} while the electrical power dissipated in the target is deduced from {6} by measuring its potential and resistance values. Finally, electrical plasma power is obtained by subtracting electrical power deposited in the target to the total power delivered by the generator as expressed in {7}.

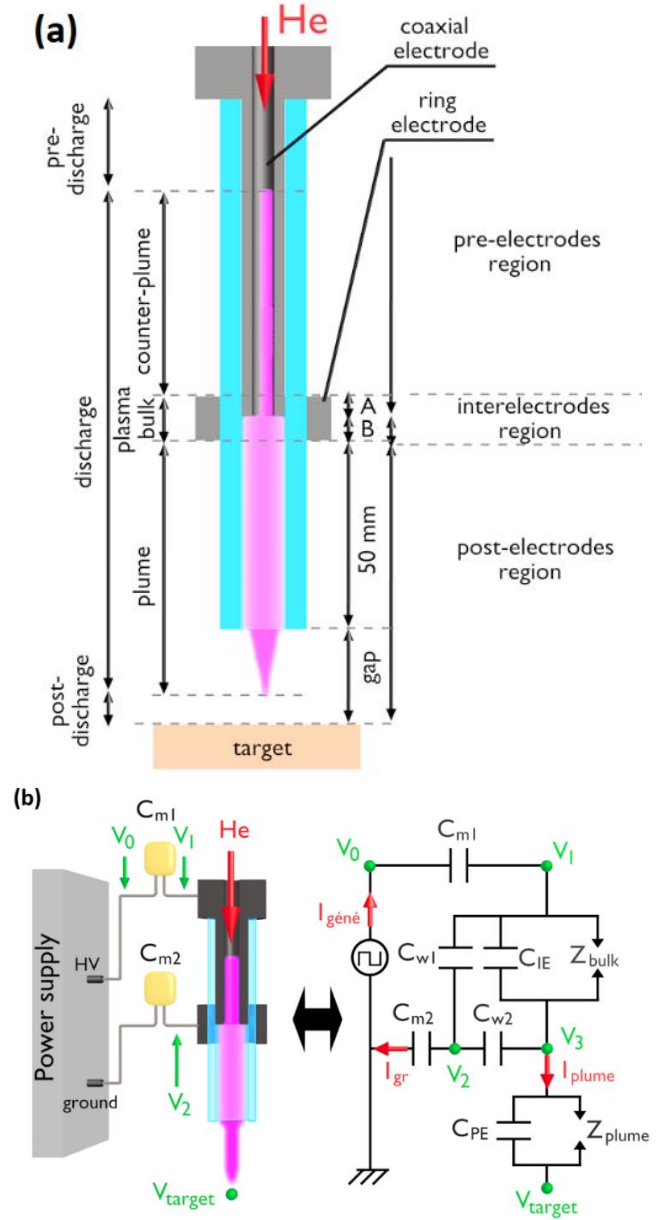

*Fig. 1 (a) Cross section view and (b) equivalent electrical circuit of the Plasma Gun (PG) connected to power supply*

$$I_{gene}(t) = \frac{d(Q_{gene}(t))}{dt} \quad \{1\}$$

$$I_{gene}(t) = C_{m1} \frac{d(V_0(t) - V_1(t))}{dt} \quad \{2\}$$

$$I_{gr}(t) = C_{m2} \frac{d(V_2(t))}{dt} \quad \{3\}$$

$$I_{plume}(t) = I_{gene}(t) - I_{gr}(t) \quad \{4\}$$

$$P_{gene} = \frac{1}{T} \int Q_{gene}.dU_{gene} = \frac{C_{m1}}{T} \int (V_0(t) - V_1(t)).dV_0 \quad \{5\}$$

$$P_{target} = \frac{1}{T} \int_0^T V_{target}(t) \times I_{target}(t).dt = \frac{1}{T} \int_0^T \frac{V_{target}^2(t)}{R_h}.dt \quad \{6\}$$

$$P_{plasma} = P_{gene} - P_{target} \quad \{7\}$$

# III. From the human body to its equivalent electrical circuit: the EEHB target

From an electrical point of view, any biological body can be represented by a circuit including resistors and capacitors. Their values depend on the properties, the number and the dimensions of the tissues and organs composing this body, and therefore on the type of the considered *in vivo* species (e.g. human, pig, mouse, etc.). Here, we propose to mimic the case of a human body with an equivalent electrical target composed of a resistor ($R_h$) in parallel with a capacitor ($C_h$) [Miklavcic et al 2006]. This equivalent electrical human body (EEHB) target is proposed to understand how a patient's body exposed to the plasma gun can influence the plasma electrical properties. Moreover, it will help to delimit an experimental operating window where electrical parameters (e.g. duty cycle, repetition frequency, etc.) are accurately defined to anticipate any electrical hazards. In the framework of *in vivo* campaigns carried out on murine models, it could also be implemented (i.e. with $R_h$ and $C_h$ values specific to mice) to limit the number of experimental conditions and therefore the number of mice to be sacrificed.

To determine the $R_h$ and $C_h$ values of the EEHB target, simulations are carried out using the Sim4Life software (Zurich MedTech, Zürich, Switzerland). This quasi-static electromagnetic solver is applied to a computable human phantom, i.e. a human body model obtained from magnetic resonance imaging (MRI) scan. The phantom selected for this work is referenced as DUKE and presents the following features: Male, 34 years, 1.77 m, 70.2 kg, BMI 22.4 kg/m² (Gosselin et al 2014). The corresponding 3-dimensional reconstitution of this human body is sketched in Figure 2.a, including 300 tissues, organs, bones as well as blood vessels. All these organization levels are subdivided into voxels (three-dimensional pixels) with a resolution as high as 0.5 x 0.5 x 0.5 mm³.

In this simulation, the PG-patient interaction is only considered on the point of electrical view. If additional radiative, chemical, flowing and thermal properties belong the plasma, their actions operate on longer time scales and/or are likely to induce less important effects/damages. Here, the electrodes of the PG are placed on DUKE's body as follows: (i) the counter-electrode is grounded (0 V), represented by two blocks placed under DUKE's feet, making a surface contact of 240 cm² and (ii) the high voltage electrode is applied on the right forearm where it delivers a sinusoidal voltage defined as $V_{target}(t)=A.sin(2\pi.f.t)$ where f=1 Hz-500 kHz and A=1-10 kV. This electrode has a cylindrical shape and allows a contact area comprised on the 10-800 mm² range. For a given frequency, the software assigns to each voxel of the Duke phantom a unique pair of values: one for the electrical conductivity and the other for the relative permittivity. Then, the software resolves Maxwell equations in every voxel where it determines the corresponding spatial distribution of the current density $\vec{J}$. The simulation result is shown in Figure 2b for f=1 kHz and A=1 kV. Instead of displaying $\vec{J}$, the result is given as a gain expressed in dB and defined as G=log(J/$J_{max}$) where J=f(x,y,z) and $J_{max}$= 1000 A.m$^{-2}$ is imposed by the software. In this simulation, the electrical current through head and left arm is negligible. Also, bones and lungs appear as very poor electrical conductors since very low gains are obtained. In the same time, $\vec{J}$ appears much stronger at knees and elbow where the same current value passes through a much narrower relative section of conductive tissues compared with bones.

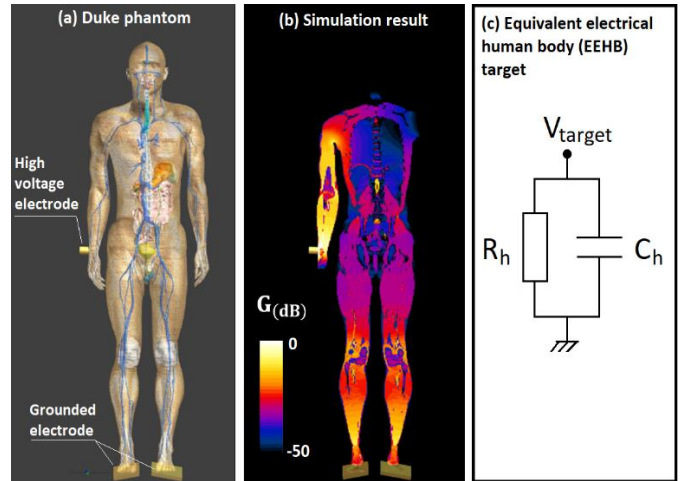

*Fig. 2 : (a) Duke phantom, (b) Simulation result showing the gain of the current density when Duke is exposed to plasma (sinusoidal voltage, 1 kHz, 1 kV, $S_c = 50 \; mm^2$) wrapped on the right forearm. Images are extracted from Sim4Life software (c) Equivalent electrical circuit of the human body, i.e. EEHB target.*

To replace this complex simulation with a simple experimental target, complex current is calculated along cross sections of the DUKE phantom by summing all the current densities $J_{vox}\downarrow$ per elementary voxel area $dS_{vox}$ as expressed in {8}. Finally, the complex admittance Y formula from {9} allows to deduce a simple equivalent electrical circuit with capacitor and resistor in parallel as described in Figure 2c.

$$\underline{I}_{cross} = \Delta S_{vox} \cdot \sum_{k=1}^{k_{max}} \underline{J}_{vox\downarrow,k} \quad \{8\}$$

$$\underline{Y} = \underbrace{\frac{Re[\underline{I}_{cross}]}{V_{applied}} + j \cdot \frac{Im[\underline{I}_{cross}]}{V_{applied}}}_{Simulation} = \underbrace{\frac{1}{R_h} + j\omega C_h}_{Theorical\ model} \quad \{9\}$$

The equivalent resistance and equivalent capacitance of the DUKE phantom are studied as a function of the high voltage electrode surface in contact with the phantom ($S_C$), amplitude and frequency of the target voltage, considering as tracking parameter the current density. Since the amplitude has no influence on $R_h$ and $C_h$, its value has been arbitrarily fixed to 1 kV to further simulate the variations of $S_C$ and f.

In Figure 3.a, $S_C$ is simulated for f=1 kHz, A=1 kV. Decreasing $S_C$ from 800 mm² to 10 mm² leads to an exponential increase in $R_h$ from 992 Ω to 1526 Ω and decrease in $C_h$ from 2.24 nF to 2.07 nF. At 1 kHz, the skin electrical conductivity presents a typical value as low as 0.20-0.66 mS/m while it is 100 times higher for bone and 1000 times higher for muscle (Bernard 2007). Accordingly, the highest electrical resistivity in human body is the one of the skin. Increasing $S_C$ reduces the skin resistance in agreement with the Pouillet's law (R=$\rho$.e/$S_C$) which connects resistance and resistivity. Consistently, a drastic increase of the skin resistance (and therefore of $R_h$) would be obtained either by decreasing $S_C$ or by applying plasma or on another part of the body where skin layer is thicker. Finally, the Figure 3.a shows that increasing $S_C$ from 10 to 50 mm² induces a rapid increase in $C_h$, followed by a plateau Such trend of $C_h$ can be explained at the light of an equivalent electrical model constituted of two capacitances in series so that:

$$\frac{1}{C_h} = \frac{1}{C'} + \frac{1}{C''} \quad (X)$$

where C' represents the capacitance in the vicinity of the electrode, i.e. the capacitance of a skin region layer and (ii) C" represents the capacitance of the rest of the body. For remind, the $\varepsilon_r$ values are almost the same whatever the biological constituents of the body. Therefore:

- For small electrode surfaces, i.e. comprised between 10 and 50 mm², we assume $C' \ll C''$ so that $C_h \approx C'$ where $C' = \varepsilon_0 \varepsilon_r \frac{S_{electrode}}{e_{skin}} \propto S_{electrode}$. Therefore, increasing $S_{electrode}$ on the 10-50 mm² range induces an increase in C' and therefore in $C_h$.
- For electrode surfaces larger than 50 mm², the high value of $S_{electrode}$ is responsible for a large value of C', so that $C' \gg C''$. According to the relation (X), it turns out that $C_h \approx C''$ and therefore remains more or less constant since the capacitance of the body remains unchanged.

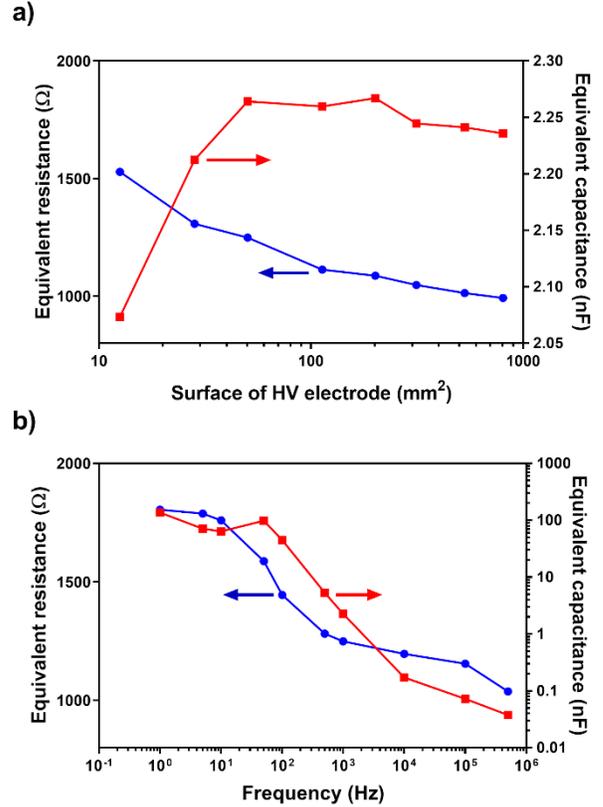

*Fig. 3: Resistance and capacitance values of the EEHB target versus (a) HV electrode surface $S_C$ for f=1 kHz, A= 1kV, (b) frequency for A=1 kV and $S_C$=50 mm².*

In figure 3.b, $R_h$ and $C_h$ are plotted vs frequency for $S_C$=50 mm² (radius 4 mm) and A=1 kV. These 2 parameters are frequency-dependant so that 3 ranges can be distinguished: (1-10Hz) no influence of frequency, (10 Hz-10 kHz) strong decrease of $R_h$ falling from 1760 Ω to 1200 Ω and of $C_h$ from 100 nF to 0.2 nF and (>10 kHz) a quasi-linear decrease of $R_h$ and $C_h$. Increasing f induces a rise in electrical conductivity of the tissues (i.e. a decrease in their electrical resistance) as well as a decay of their relative permittivities (i.e. a decrease in capacitance) [Bernard2007]. Two effects can explain such trends:

- (i) The α relaxation, which results from ionic diffusion at the scale of cells membrane or other biological structures and which can lead to the formation of electric dipoles.
- (ii) The β relaxation, a type of Maxwell Wagner relaxation, i.e. an interface relaxation that appears in any complex medium composed of sub-media showing specific electrical properties.

For the following investigations, the EEHB target has been calibrated at f=30kHz (corresponding to the repetition frequency utilized upon *in vivo* treatments) and for $S_C$ corresponding to the inner area of the PG tube, i.e. 12.6 mm². Resistance and capacitance values have been estimated accordingly, leading to the choice of the following components $R_h$=1500 Ω (Vishay, 50W) and $C_h$=100 pF (Murata, 6.3 kV DC).

# IV. Plasma gun / EEHB target interaction: measuring & correlating electrical plasma power, electrical deposited power & currents profiles

Two shortcuts are commonly taken when it comes to evaluate the plasma power of a plasma source interacting with a target: (i) assuming that the electrical charge is the same at every point of the electrical circuit, (ii) confounding plasma power with power measured between the two electrodes, regardless of the target properties. In that framework, the plasma power is commonly estimated by plotting the electrical charge at the grounded electrode ($C_{m2}$) as a function of the plasma voltage ($V_{pl,rough} \simeq V_0 - V_2 \simeq V_0$) and by assessing the area of the resulting Lissajous curve. As an alternative, the most accurate approach consists into evaluating the electrical plasma power ($P_{pl,a}$) dissipated in the interelectrode and post-electrode regions as sketched in Figure 1a. For this, (i) $C_{m1}$ must be inserted as sketched in Figure 1b to evaluate the total charge in the circuit, (ii) $P_{gene}$ must be calculated with {5} and (iii) $P_{pl,a}$ deduced from $P_{target}$ and $P_{gene}$ using {7}.

As shown in Figure 4b for an EEHB target 10 mm away from the PG supplied with $V_0$=9 kV, the rough approach leads to $P_{Pl,rough}$=11.0 W. A strong discrepancy appears if one measures the electrical power dissipated in the target ($P_{target}$) since a value as high as 61 W is measured. As an alternative, using the most accurate approach gives $P_{gene}$=184 W and an accurate value of the plasma power as high as $P_{Pl,a}$=123 W. Such values, quite high at first sight, are consistent with the given conditions where plasma behaves as a transient arc. On the contrary, for $V_0$=8 kV, a non thermal plasma is obtained with $P_{gene}$ close to 40 W, always using the most accurate method. These results clearly indicate that $P_{plasma}$ is always underestimated when electrical charge is calculated using $C_{m2}$ instead of $C_{m1}$.

To understand how $P_{plasma}$ can reach such elevated values (61 W), time profiles of $I_{plume}$ and plasma voltage are plotted versus time considering the two aforementioned conditions ($V_0$ = 8 and 9 kV). In all cases, $I_{plume}$ is determined by equation {4} and is the sum of a capacitive current (from $C_{PE}$) and a discharge current (from $Z_{plume}$). For $V_0$=8 kV, the two peaks partially overlap in time, hence giving rise to a single and slightly asymmetrical peak appearing twice per repetition period, i.e. at the rising and falling edges of the voltage pulse as observed in Figure 5.

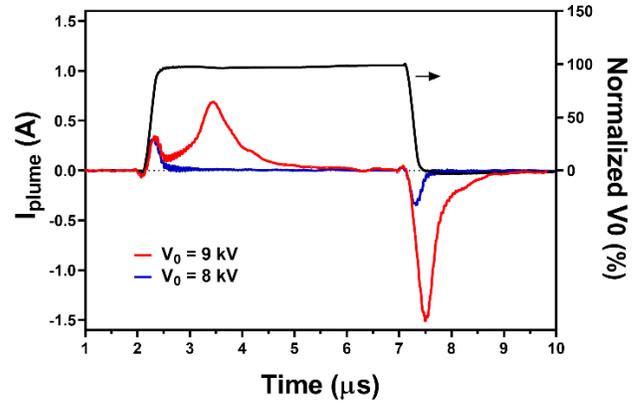

*Fig. 5: Influence of targets on instantaneous current in Plasma Gun's plume. Experimental conditions: Voltage 9kV, helium flow 1000 sccm, frequency 30 kHz and duty cycle 14 %.*

For $V_0$=9 kV, the profile of $I_{plume}$ is quite different. During the voltage state (9 kV) a second positive current peak appears 1.4 µs after the first peak. If the first peak is similar to the one at 8 kV in terms of amplitude, triggering on the voltage rising edge and rising/decaying durations, the second peak appears much larger in amplitude and duration. At 7.2 µs a single but high negative current peak appears at the failing edge of the low voltage state (0 V). If its rising slope is the same as for $V_0$=8 kV, its maximum value is much higher, rising -1.8 A and its duration appears twice longer.

Two plasma propagation mechanisms can be identified depending on the applied voltage $V_0$ :

- For $V_0$<8.5 kV: a single peak of current - narrow and symmetric - is obtained per high/low voltage state. Accordingly, plasma is assumed to behave as an ionization wave with a local electric field of high magnitude in the ionization front, and a tail whose length is subject to two interpretations. The first one relies on assuming

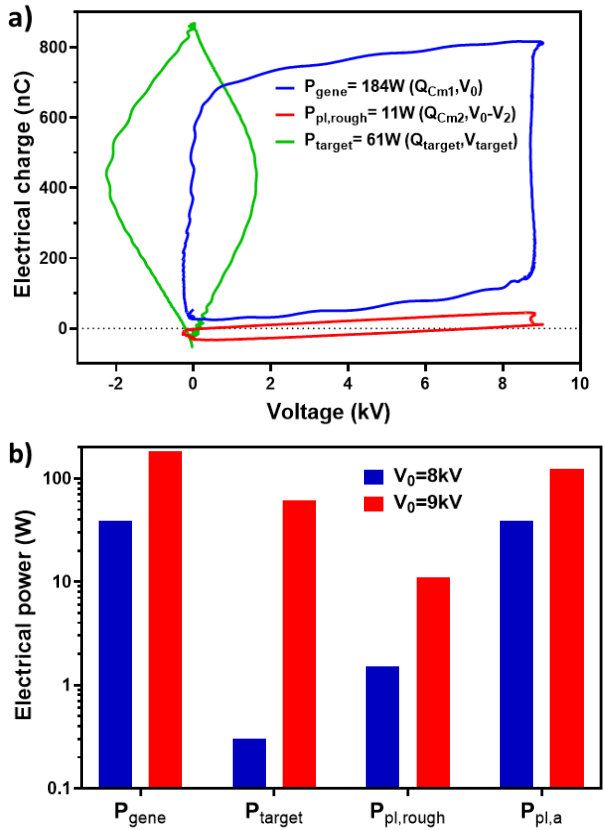

*Fig. 4. (a) Lissajous curves of the PG-EEHB configuration for $V_{pl}$=9 kV, (b) electrical powers of the generator, target and plasma estimated for $V_0$=8 and 9 kV. Conditions: gap=10 mm, f=30 kHz, Duty cycle=14 %*

a tail as long as the propagation length of the plume although its ionization magnitude is too low to be electrically detected along the interelectrode distance. The wave front remains connected to the inner HV electrode through a low-emissive channel. Such bridging lasts as long as the voltage between HV electrode and the floating conductive target becomes too low to keep open the ionized channel. The second interpretation relies on the assumption that the tail may be shorter than the propagation length of the plume. In that latter case the concept of plasma bullet might be utilized, although it induces the absence of connexion between the ionization wave and the electrode.

- Once an electrical contact is established between plasma and target, a PAPS-to-arc transition can be obtained under specific conditions:

(i) If $C_{m1}$ is part of the circuit, the equivalent electrical model is a RC series circuit including $C_{m1}$, $Z_{plasma}$ (plasma impedance) and $Z_{target}$ (target impedance). In that case, a transient arc is formed above 8.5 kV and is characterized in Figure 5 by a second and wide current peak obtained upon the positive voltage state. The profile of this second peak corresponds to the waveform of a RC series circuit.

(ii) If $C_{m1}$ is not part of the circuit the equivalent electrical model is limited to $Z_{plasma}$ in series with $Z_{target}$. The plasma bridges the HV electrode to the target so that the PAPS-to-arc transition is still effective although on a more intense way: instead of obtaining two discreet current peaks, a current waveform similar to that of the pulse generator voltage would be obtained (duty cycle of 14 %). The reason is that the absence of $C_{m1}$ permits the determination of the target potential using a voltage divider law.

# V. Conclusion

Simulations performed with the Sim4life software has enabled us to engineer and characterize a new model of target mimicking the electrical response of human body exposed to the plasma gun. After assessing current densities in each voxel of the DUKE phantom, complex admittances have been computed on various cross sections of the phantom and return a simple equivalent electrical circuit with resistor and capacitor in parallel. The values of $R_h$=1500 Ω and $C_h$=100 pF have been determined.

Experimental works have been carried out to measure electrical power at relevant points of PG and EEHB equivalent electrical circuits. We strongly underline that, for the sake of accuracy, plasma power measurements should be performed by subtracting the target power to the generator power. Usual Lissajous method can lead to significantly underestimated plasma powers because of the electrical charge which is not constant in a three electrodes configuration like the PG/EEHB circuit. Besides, we underline that if the plasma gun has shown therapeutic effects on murine models, great care should be brought in the framework of translational research performed on human patients. In that case, there exists a critical value ($V_0$≈8.5 kV in our setup) beyond which electrical hazards probability becomes important: below this value, pulsed atmospheric plasma streams are emitted towards the EEHB target while beyond this threshold, plasma reaches thermal equilibrium under the form of a transient electrical arc.

For further rigorous practice at hospital, the plasma source and the patient cannot be envisioned separately: on the point of electrical view, they must be considered as a unique and unified system. Before exposing a patient to the plasma gun, a preliminary step appears as a mandatory so as to calibrate the interactive PG-patient system. In practice, this means accurately assessing the values of $R_h$ and $C_h$ which depend on one hand on tissues biophysical parameters (relative permittivities, electrical resistivity, thicknesses, …) and on on the other hand on physical parameters specific to the plasma gun (repetition frequency, amplitude, gap, electrode surfaces). This preliminary step is crucial to define an experimental operating window to overcome the electrical risks, in this case the formation of arcs likely to burn tissues or electrify the patient.

For the future, it seems appropriate to (i) upgrade the equivalent electrical circuit of the PG by a finer modeling of $Z_{bulk}$ and $Z_{plume}$, as suggested in the works of Fang et al [Fang2016], (ii) develop new equivalent electrical models of flagship plasma sources (e.g. kinpen, plasma gun), (iii) Create $R_h$ and $C_h$ calibration tables for living models presenting an interest for laboratory research (e.g. mice with/without tumor) and translational research (human differing in age, weight, sex, …).

# VI. Acknowledgements

This work has been achieved within the LABEX Plas@Par fundings and has been supported by grants from Région Ile-de-France (Sesame, Ref. 16016309). This work is partly supported by French network GDR 2025 HAPPYBIO.

# VII. References

*(Bernard 2007)* Bernard L 2007 *Caractérisation électrique des tissus biologiques et calcul des phénomènes induits dans le corps humain par des champs électromagnétiques de fréquence inférieure au GHz. Ecole Centrale de Lyon ; Universidade federal de Minas Gerais. PhD thesis.*

*(Brullé et al 2012)* Brullé L, Vandamme M, riès D, Martel E, Robert E, Lerondel S, Trichet V, Richard S, Pouvesle J-M, Le Pape A 2012 *Effects of a non thermal plasma traitement alone or in combinations with gemcitabine in a MIA PaCa2-luc orthotopic pancreatic carcinoma model Plos One* **7** *e52653*

*(Darny et al 2017)* Darny T, Pouvesle J-M, Puech V, Douat C, Dozias S, Robert E 2017 *Analysis of conductive target influence in plasma jet experiments through helium metastable and electric field measurements Plasma Sources Sci. Technol.* **26** *045008*

*(Fang et al 2016)* Fang Z, Shao T, Yang J, Zhang C 2016 *Discharge processes and an electrical model of atmospheric pressure plasma jets in argon Eur. Phys. J. D* **70** *60437-4*

*(Gosselin et al 2014) Gosselin M-C et al 2014 Development of a new generation of high-resolution anatomical models for medical device evaluation: the Virtual Population 3.0 Phys. Med. Biol.* **59** *5287-5303*

*(Guaitella and Sobota 2015) Guaitella O, Sobota A 2015 The impingement of a kHz helium atmospheric pressure jet on a dielectric surface J. Phys. D: Appl. Phys.* **48** *255202*

*(Kone et al 2017) Kone A, Sainct F P, Muja C, Caillier B, Guillot P 2017 Investigation of the interaction between a helium jet and conductive (Metal) / non-conductive (Dielectric) target Plasma Med.* **7** *4*

*(Kostov et al 2009) Kostov K G, Honda R Y, Alves L M S, Kayama M E 2009 Characteristics of dielectric barrier discharge reactor for material treatment Braz. J. Phys.* **39** *322-325*

*(Koumbourlis 2002) Koumbourlis A C 2002 Electrical injuries Crit Care Med* **30** *11*

*(Laroussi 2015) Laroussi M 2015 Low-Temperature Plasma Jet for Biomedical Applications: A Review IEEE Trans. Plasma Sci.* **43** *703-712*

*(Miklavcic et al 2006) Miklavcic D, Pavselj N, Hart F X 2006 Electrical properties of tissues Wiley Encyclopedia of Biomedical Engineering John Wiley & Sons*

*(Norberg et al 2015) Norberg S A, Johnsen E, Kushner M J 2015 Helium atmospheric pressure plasma jets touching dielectric and metal surfaces J. Appl. Phys.* **118** *013301*

*(Park et al 2018) Park I, Kim S-Y, Bae M-k, Lee Y, Song Y, Lim T H, Chung K 2018 Comparison of AC plasma jets between dielectric barrier discharge and surface barrier discharge Clin. Plasma Med.* **9** *7*

*(Robert et al 2012) Robert E, Sarron V, Ries D, Dozias S, Vandamme M, Pouvesle J-M 2012 Characterization of pulsed atmospheric-pressure plasma streams (PAPS) generated by a plasma gun Plasma Sources Sci. Technol.* **21** *034017*

*(Sarron et al 2011) Sarron V, Robert E, Dozias S, Ries D, Vandamme M, Pouvesle J-M 2011 Propagation of two symmetrical pulsed atmospheric plasma streams generated by a pulsed plasma gun 20th International Symposium on Plasma Chemistry*

*(Slutsker et al 2017) Slutsker Y Z, Semenov V E, Krasik Y E, Ryzhkov A, Felsteiner J, Binenbaum Y, Gil Z, Schtricman R, Cohen J T 2017 Electrical model of cold atmospheric plasma gun Phys. Plasmas* **24** *103510*

*(Weltmann and Woedtke 2017) Weltmann K-D, von Woedtke Th 2017 Plasma medicine-current state of research and medical application Plasma Phys. Control. Fusion* **59** *014031*

*(Winter et al 2015) Winter J, Brandenburg R, Weltmann K-D 2015 Atmospheric pressure plasma jets: an overview of devices and new directions Plasma Sources Sci. Technol.* **24** *064001*